\documentclass[prd,aps,nofootinbib,floatfix,11pt]{revtex4}
\usepackage{amsmath,graphicx,epsfig,amssymb,dsfont,mathtools,}
\usepackage[usenames]{color}
\usepackage{ulem} 
\usepackage{bigstrut}
\usepackage{slashed}
\usepackage{multirow}
\usepackage{subfigure}
\usepackage{threeparttable}

\allowdisplaybreaks

\begin{document}
\title{QCD sum rules study on weak decays $\Omega_c^0\to\Omega^- (\pi^+, \rho^+, l^+\nu_l)$}

\author{Yu-Ji Shi$^{1}$~\footnote{Email:shiyuji@ecust.edu.cn (Corresponding author)},
  Jun Zeng~$^{2,3}$~\footnote{Email:zengj@alu.cqu.edu.cn (Corresponding author)}}
\affiliation{$^1$ School of Physics, East China University of Science and Technology, Shanghai 200237, China\\
$^{2}$ College of Physics and Electronic Engineering, Hainan Normal University, Haikou, Hainan Province, 571158, China\\
$^{3}$ INPAC, Key Laboratory for Particle Astrophysics and Cosmology (MOE),  Shanghai Key Laboratory for Particle Physics and Cosmology, School of Physics and Astronomy, Shanghai Jiao Tong University, Shanghai 200240, China}

\begin{abstract}
We study the weak decays of the charmed baryon $\Omega_c^0$ to the $\Omega^-$ baryon within the framework of QCD sum rules. A three-point correlation function is defined for calculating eight form factors governing the $\Omega_c^0 \to \Omega^-$ transition. The cutting rules are employed to extract the double imaginary parts of the correlation functions, enabling their expression in the form of dispersive integrals. These form factors are then used to determine the branching fractions for the hadronic decays $\Omega_c^0 \to \Omega^- \pi^+$ and $\Omega_c^0 \to \Omega^- \rho^+$, as well as the semi-leptonic decay $\Omega_c^0 \to \Omega^- \ell^+ \nu_\ell$. Our results are compared with existing theoretical predictions and experimental data, providing a comprehensive analysis of $\Omega_c^0$ decays and offering valuable insights into the dynamics of charmed baryon decays.
\end{abstract}
\maketitle

\section{Introduction}
The decays of charmed baryons provide a unique opportunity to probe the interplay between strong and weak interactions, as well as exploring the non-perturbative dynamics of QCD. Among the charmed baryons, the anti-triplet states: $\Lambda_c^+, \Xi_c^0, \Xi_c^+$ and the sextet states: $\Sigma_c^{(0,+,++)}, \Xi_c^{'(0,+)}, \Omega_c^0$ have been the subject of extensive experimental and theoretical investigations. The $\Sigma_c$ and $\Xi'_c$ baryons predominantly decay via strong and electromagnetic decays, while the anti-triplet and $\Omega_c^0$ undergo  weak interactions. Recent experiments have significantly improved our understanding of the weak decays of $\Lambda_c$ and $\Xi_c$ baryons, with benchmark modes such as $\Lambda_c \to p K^- \pi^+$ \cite{Belle:2013jfq,BESIII:2015bjk} and $\Xi_c^0 \to \Xi^- \pi^+$ \cite{Belle:2018kzz} being precisely measured. These results have provided critical insights into the underlying mechanisms of charmed baryon decays.

In comparison, the experimental exploration of weak decays involving the $\Omega_c^0$ baryon remains relatively limited. While a direct measurement of the benchmark decay mode $\Omega_c^0 \to \Omega^- \pi^+$ is still lacking, experimental studies have instead concentrated on determining relative branching fractions, expressed as
\begin{equation}
R(X) = \frac{\mathcal{B}(\Omega_c^0 \to X)}{\mathcal{B}(\Omega_c^0 \to \Omega^- \pi^+)},
\end{equation}
which are experimentally more tractable. Recent results from LHCb \cite{LHCb:2023fvd}, ALICE \cite{ALICE:2024xjt}, and Belle \cite{Belle:2022yaq,Belle:2021dgc,Belle:2017szm} have provided crucial data on $R(X)$ for decays such as $\Omega_c^0 \to \Omega^- K^+$, $\Xi^- \pi^+$, and $\Omega^- e^+ \nu_e$. These measurements emphasize the necessity of precise theoretical predictions for the absolute branching fraction of $\Omega_c^0 \to \Omega^- \pi^+$, which are vital for interpreting experimental observations.

The $\Omega_c^0 \to \Omega^- \pi^+$ decay is particularly interesting from a theoretical perspective due to its simple kinematic and dynamic structure. This decay is governed solely by factorizable processes, with non-perturbative effects entirely encapsulated in the baryon transition form factors. This simplicity makes it an ideal candidate for theoretical studies, as it allows for a clearer understanding of the underlying QCD dynamics. Furthermore, the decay $\Omega_c^0 \to \Omega^- \pi^+$ exhibits a topological structure analogous to other modes, such as $\Omega_c^0 \to \Omega^- \rho^+$ and $\Omega_c^0 \to \Omega^- \ell^+ \nu_\ell$. This similarity implies that a unified theoretical description of these decays is achievable, contingent upon an accurate determination of the $\Omega_c^0 \to \Omega^-$ transition form factors.

Theoretical studies of $\Omega_c$ decays have a long history, with early work dating back to the 1990s \cite{Xu:1992sw,Cheng:1996cs}. However, theoretical interest in this area diminished until renewed experimental progress spurred further investigations. Recent studies have adopted a range of approaches, including the light-front quark model \cite{Hsiao:2020gtc,Zhao:2018zcb}, the constituent quark model \cite{Gutsche:2018utw,Wang:2022zja}, non-relativistic quark model \cite{Pervin:2006ie,Zeng:2024yiv}, light-cone sum rules \cite{Aliev:2022gxi,Aliev:2018uby}, hybrid fitting techniques \cite{Liu:2023dvg,Hsiao:2023mvw}, and  topological diagrammatic approach combined with the MIT bag model  \cite{Cheng:2018hwl,Zou:2019kzq,Cheng:2020wmk,Meng:2020euv,Cheng:2022kea,Wang:2023uea,Luo:2023vbx}. 
In this work, we study $\Omega_c^0 \to \Omega^-$ decays by employing the QCD sum rules (QCDSR), a model-independent and QCD-based framework, and calculate the form factors governing the $\Omega_c^0 \to \Omega^-$ transition. Using these form factors, we compute the branching fractions for the benchmark mode $\Omega_c^0 \to \Omega^- \pi^+$ as well as other related decay modes, such as $\Omega_c^0 \to \Omega^- \rho^+$ and $\Omega_c^0 \to \Omega^- \ell^+ \nu_\ell$. Our results are compared with existing theoretical predictions and experimental data, providing a comprehensive and robust analysis of $\Omega_c^0$ decays. This study not only enhances our understanding of the $\Omega_c^0$ decays within QCD but also establishes a reliable foundation for future theoretical and experimental studies.

The paper is organized as follows: In Section~\ref{sec:correQCDSR}, we introduce the basic framework of QCDSR and define the correlation function used for extracting the $\Omega_c^0 \to \Omega^-$ form factors. In Section~\ref{sec:hadronLevel} and \ref{sec:QCDLevel}, we respectively introduce the hadron level and QCD level calculation for the correlation function. In Section \ref{sec:numericalResult}, we provide the numerical results for the form factors and branching fractions, along with a detailed comparison with previous studies. Section~\ref{sec:conclusion} is a summary of this work.

\section{Correlation function in QCDSR}\label{sec:correQCDSR}

The effective weak Hamiltonians for the hadronic and semileptonic $\Omega_c\to\Omega$ decays are  
\begin{align}
{\cal H}_H&=\frac{G_F}{\sqrt{2}}V_{cs}^*V_{ud}\left[c_1(\bar u d)(\bar s c)+c_2(\bar s d)(\bar u c)\right],\nonumber\\
{\cal H}_L&=\frac{G_F}{\sqrt{2}}V_{cs}^*(\bar s c)({\bar u}_{\nu} \nu_l),
\end{align}
where $G_F$ is the Fermi constant, $V_{cs}$ and $V_{ud}$ are the Cabibbo–Kobayashi–Maskawa (CKM) matrix elements, $c_1$ and $c_2$ are the Wilson coefficients, and the notation $({\bar q}_1q_2) = {\bar q}_1\gamma_{\mu}(1-\gamma_5)q_2$ is adopted. As demonstrated by factorization in Ref.~\cite{Hsiao:2019ann}, the transition matrix elements induced by $(\bar s d)(\bar u c)$ are equivalent to those induced by $(\bar u d)(\bar s c)/N_c$. Consequently, ${\cal H}_H$ can be simplified as
 \begin{align}
{\cal H}_H&=\frac{G_F}{\sqrt{2}}V_{cs}^*V_{ud} a_1(\bar u d)(\bar s c),\label{eq:EffHamilton}
\end{align}
with $a_1=c_1+c_2/N_c$. The transition matrix element of $\Omega_c\to\Omega$ can be parameterized by 8 form factors:
\begin{align}
&\langle \Omega(p^{\prime}, s^{\prime})|\bar s \gamma^{\mu}(1-\gamma_5)c|\Omega_c(p, s)\rangle\nonumber\\
=&{\bar u}_{\alpha}(p^{\prime}, s^{\prime})\left[\frac{p^{\alpha}}{M_1}\left(\gamma^{\mu}f_1(q^2)+\frac{p^{\mu}}{M_1}f_2(q^2)+\frac{p^{\prime\mu}}{M_2}f_3(q^2)\right)+g^{\alpha\mu}f_4(q^2)\right]\gamma_5 u(p,s)\nonumber\\
&-{\bar u}_{\alpha}(p^{\prime}, s^{\prime})\left[\frac{p^{\alpha}}{M_1}\left(\gamma^{\mu}g_1(q^2)+\frac{p^{\mu}}{M_1}g_2(q^2)+\frac{p^{\prime\mu}}{M_2}g_3(q^2)\right)+g^{\alpha\mu}g_4(q^2)\right] u(p,s),\label{eq:formfactorDef}
\end{align}
where $q=p-p^{\prime}$, $M_1, M_2$ are the masses of $\Omega_c, \Omega$. $f_{1,2,3,4}$ and $g_{1,2,3,4}$ are the form factors for the transitions induced by vector and axial-vector currents, respectively.

Within the framework of QCDSR, we define a three-point correlation function to calculate the 8 form factors for the $\Omega_c \to \Omega$ transition, as defined in Eq.~(\ref{eq:formfactorDef}):
\begin{align}
\Pi_{\alpha\mu}(p_2,p_1)=i^2 \int d^4 x d^4 y\ e^{ip_2\cdot y}e^{-ip_1\cdot x}\langle0|T\{J_{\alpha}^{\Omega}(y)J_{\mu}(0){\bar J}^{\Omega_c}(x)\}|0\rangle,\label{eq:correFunc}
\end{align}
where $J_{\alpha}^{\Omega}$ and $J^{\Omega_c}$ are the interpolating currents for $\Omega$ and $\Omega_c$ baryons\cite{Aliev:2022gxi,Wang:2009cr,Shi:2019hbf}:
\begin{align}
J_{\alpha}^{\Omega}&=\epsilon^{ijk}(s_i^TC\gamma_{\alpha}s_j)s_k,\nonumber\\
J^{\Omega_c}&=\epsilon^{ijk}(s_i^TC\gamma^{\beta}s_j)\gamma_{\beta}\gamma_5 c_k,
\end{align}
$J_{\mu}={\bar s}\gamma_{\mu}(1-\gamma_5)s$ is the $V-A$ current. Note that the correlation function in Eq.~(\ref{eq:correFunc}) contains two Lorentz indices, resulting in dependence on numerous spinor and Lorentz structures, most of which are redundant for extracting the form factors. To simplify the correlation function, we perform the following four projections to eliminate the Lorentz indices:
\begin{align}
&\Pi_{i}(p_2,p_1)={\cal P}_i^{\alpha\mu}\Pi_{\alpha\mu}(p_2,p_1),\nonumber\\
&{\cal P}_i^{\alpha\mu}=\{p_1^{\alpha}p_2^{\mu},\  p_1^{\alpha}p_1^{\mu},\ M_1^2 g^{\alpha\mu},\ M_1 p_1^{\alpha}\gamma^{\mu}\}.
\end{align}
Each $\Pi_i(p_2, p_1)$ can be expanded in terms of 8 independent Dirac structures:
\begin{align}
&\Pi_i(p_2,p_1)=\sum_{k=1}^8 C_i^k(p_2,p_1) N_k,\nonumber\\
&N_k=\{\slashed p_2\slashed p_1\gamma_5,\ \slashed p_2\gamma_5,\ \slashed p_1\gamma_5,\ \gamma_5,\ \slashed p_2\slashed p_1,\ \slashed p_2,\ \slashed p_1,\ 1\}.
\end{align}

The four correlation functions $\Pi_i$ are calculated separately at both the hadron and QCD levels. These results are then equated using the principle of quark-hadron duality. After applying the Borel transformation, the final sum rules for extracting the form factors are obtained, which can be expressed as
\begin{align}
\Pi_i^{H}(p_2,p_1)&=\sum_{k=1}^8 C_i^{k, H}(p_2,p_1) N_k,\nonumber\\
\Pi_i^{\rm QCD}(p_2,p_1)&=\sum_{k=1}^8 C_i^{k, \rm QCD}(p_2,p_1) N_k^{\rm QCD},\nonumber\\
{\cal B}_{T_1^2, T_2^{2}}\left[C_i^{k, H}(p_2,p_1)\right]&={\cal B}_{T_1^2, T_2^{2}}\left[C_i^{k, \rm QCD}(p_2,p_1)\right], \label{eq:BorelSumEq}
\end{align}
where ${\cal B}_{T^2, T_2^2}$ denotes the double Borel transformation applied to $p_1^2$ and $p_2^2$, with $T^2$ and $T^{\prime 2}$ representing the corresponding Borel parameters. Since $C_i^{k, H}(p_2,p_1)$ depends on the eight form factors, Eq.~(\ref{eq:BorelSumEq}) can, in principle, be utilized to determine all of them. However, Eq.~(\ref{eq:BorelSumEq}) yields 32 equations, significantly exceeding the number of form factors. To ensure a unique solution, we select $k=1$ and $k=5$ in Eq.~(\ref{eq:BorelSumEq}) to solve for $f_i(q^2)$ and $g_i(q^2)$, respectively. This selection is justified by the fact that $C_i^{1, \rm QCD}(p_2,p_1)$ and $C_i^{5, \rm QCD}(p_2,p_1)$, corresponding to the coefficients of the highest-dimensional Dirac structures $\slashed p_2\slashed p_1\gamma_5$ and $\slashed p_2\slashed p_1$, are suppressed by the highest power of $1/q^2$. This suppression is crucial for ensuring the convergence of the operator product expansion (OPE) at the QCD level.

\section{Hadron level calculation}\label{sec:hadronLevel}

At the hadron level, we insert complete sets of intermediate states with the same quantum numbers as $\Omega_c$ and $\Omega$ into the correlation function $\Pi_{i}$:
\begin{align}
\Pi_{i}^{H}(p_{2},p_{1})  =&i^{2}\int d^{4}xd^{4}y\ e^{ip_{2}\cdot y}e^{-ip_{1}\cdot x}\sum_{s,s^\prime}\int\frac{d^{4}p^{\prime}}{(2\pi)^{4}}\frac{d^{4}p}{(2\pi)^{4}}\frac{i}{p_2^{2}-M_2^{2}}\frac{i}{p_1^{2}-M_1^{2}}\nonumber\\
 & \times{\cal P}_{i}^{\alpha\mu}\langle0|T\{J_{\alpha}^{\Omega}(y)|\Omega(p^{\prime},s^{\prime})\rangle\langle\Omega(p^{\prime},s^{\prime})|J_{\mu}(0)|\Omega_{c}(p,s)\rangle\langle\Omega_{c}(p,s)|{\bar{J}}^{\Omega_{c}}(x)\}|0\rangle+\cdots,\label{eq:insertstates}
\end{align}
where we have explicitly isolated the term corresponding to the lowest states, referred to as the pole contribution, while the ellipsis represents contributions from excited states and the continuum spectrum. The first and the last matrix elements above can be expressed by the decay constants of $\Omega$ and $\Omega_c$, which are defined as
\begin{align}
\langle0|J_{\alpha}^{\Omega}(0)|\Omega(p^{\prime},s^{\prime})\rangle&=f_{\Omega}u_{\alpha}(p^{\prime},s^{\prime}),\nonumber\\
\langle0|J^{\Omega_c}(0)|\Omega_c(p,s)\rangle&=f_{\Omega_c}u(p,s).
\end{align}
It should be noted that, in principle, $J_{\alpha}^{\Omega}$ can couple to both spin-3/2 and spin-1/2 $\Omega$ states. However, since the ground state of $\Omega$ has spin-3/2, we retain only the spin-3/2 component in Eq.~(\ref{eq:insertstates}) and attribute the spin-1/2 $\Omega$ states to the contributions from excited states. Summing up the polarizations of $\Omega$ and $\Omega_c$, and using the spin summation formula for the spin-3/2 $\Omega$ baryon:
\begin{align}
\sum_{s} & u_{\alpha}(p,s)\bar{u}_{\beta}(p,s)=-g_{\alpha\beta}+\frac{1}{3}\gamma_{\alpha}\gamma_{\beta}+\frac{2}{3}\frac{p_{\alpha}p_{\beta}}{M_{2}^{2}}-\frac{p_{\alpha}\gamma_{\beta}-p_{\beta}\gamma_{\alpha}}{3M_{2}},
\end{align}
one arrive at
\begin{align}
C_{i}^{k,H}=&\frac{\lambda_{\Omega_{c}}\lambda_{\Omega}}{(p_{1}^{2}-M_{1}^{2})(p_{2}^{2}-M_{2}^{2})}\tilde{C}_{i}^{k,H},\\
\tilde{C}_1^{1, H}=&-\frac{\left(M_1^4-2 M_1^2
   \left(M_2^2+q^2\right)+\left(M_2^2-q^2\right)^2\right)
   \left(2 M_1 M_2 (f_1+f_3)+f_2
   \left(M_1^2+M_2^2-q^2\right)\right)}{12 M_1^2
   M_2^2},\nonumber\\
\tilde{C}_2^{1, H}=&  -\frac{\left(M_1^4-2 M_1^2
   \left(M_2^2+q^2\right)+\left(M_2^2-q^2\right)^2\right)
   \left(2 M_1 M_2 (-f_1+f_2+f_4)+f_3
   \left(M_1^2+M_2^2-q^2\right)\right)}{12 M_1
   M_2^3},\nonumber\\
\tilde{C}_3^{1, H}=&-\frac{4 M_1 M_2 \left(f_1
   \left((M_1+M_2)^2-q^2\right)-3 f_4 M_1
   M_2\right)+f_2 \left(M_1^4-2 M_1^2
   \left(M_2^2+q^2\right)+\left(M_2^2-q^2\right)^2\right)}{6 M_2^2},\nonumber\\
\tilde{C}_4^{1, H}=&-\frac{\left((M_1-M_2)^2-q^2\right)}{6 M_2^2}\Big[2 f_1
   \left(q^2-(M_1+M_2)^2\right)\nonumber\\
&+f_2
   \left((M_1+M_2)^2-q^2\right)+f_3 M_1^2+2
   f_3 M_1 M_2+f_3 M_2^2-f_3 q^2+4
   f_4 M_1 M_2\Big],\\
\tilde{C}_1^{5, H}=&-\frac{\left(M_1^4-2 M_1^2
   \left(M_2^2+q^2\right)+\left(M_2^2-q^2\right)^2\right)
   \left(2 M_1 M_2 (g_1+g_3)+g_2
   \left(M_1^2+M_2^2-q^2\right)\right)}{12 M_1^2
   M_2^2},\nonumber\\
\tilde{C}_2^{5, H}=&-\frac{\left(M_1^4-2 M_1^2
   \left(M_2^2+q^2\right)+\left(M_2^2-q^2\right)^2\right)
   \left(2 M_1 M_2 (g_1+g_2+g_4)+g_3
   \left(M_1^2+M_2^2-q^2\right)\right)}{12 M_1
   M_2^3},\nonumber\\
\tilde{C}_3^{5, H}=&\frac{4 M_1 M_2 \left(g_1
   \left(q^2-(M_1-M_2)^2\right)+3 g_4 M_1
   M_2\right)+g_2 \left(-M_1^4+2 M_1^2
   \left(M_2^2+q^2\right)-\left(M_2^2-q^2\right)^2\right)
   }{6 M_2^2},\nonumber\\
\tilde{C}_4^{5, H}=& \frac{\left((M_1+M_2)^2-q^2\right)}{6 M_2^2} \Big[2 g_1
   \left((M_1-M_2)^2-q^2\right)\nonumber\\
   &+g_2
   \left((M_1-M_2)^2-q^2\right)-g_3 M_1^2+2
   g_3 M_1 M_2-g_3 M_2^2+g_3 q^2-4
   g_4 M_1 M_2\Big].
\end{align}

In Eq.~(\ref{eq:BorelSumEq}), the correlation function at QCD level can be generally expressed in a form of double dispersion integration:
\begin{align}
C_{i}^{k,{\rm QCD}}(p_{1},p_{2})=\frac{1}{\pi^{2}}\int_{0}^{\infty}ds_{1}\int_{0}^{\infty}ds_{2}\frac{{\rm Im}^{2}C_{i}^{k,{\rm QCD}}(s_{1},s_{2})}{(s_{1}-p_{1}^{2})(s_{2}-p_{2}^{2})},\label{eq:QCDPidisper}
\end{align}
where ${\rm Im}^{2}$ denotes extracting the doubly imaginary part  of  $C_{i}^{k,{\rm QCD}}$ in the complex planes of $s_1$ and $s_2$. According to quark-hadron duality, the integration in the large $s_1$ and $s_2$ regions is canceled by the contribution from  higher states and continuous spectrum at hadron level.
Finally, from Eq.~(\ref{eq:BorelSumEq}) on can solve out the form factors as
\begin{align}
\left\{ f_{i},g_{i}\right\}  & =\frac{1}{\lambda_{\Omega_{c}}\lambda_{\Omega}}e^{M_{1}^{2}/T_{1}^{2}}e^{M_{2}^{2}/T_{2}^{2}}\frac{1}{\pi^{2}}\int_{0}^{s_{1\rm th}}ds_{1}\int_{0}^{s_{2\rm th}}ds_{2}\ e^{-s_1^{2}/T_{1}^{2}}e^{-s_2^{2}/T_{2}^{2}}\left\{ \tilde{f}_{i},\tilde{g}_{i}\right\},\label{eq:extractFF}
\end{align}
where $s_{1\rm th}$ and $s_{2\rm th}$ are the threshold parameters that represent the starting points of the higher excited state spectra above $\Omega_c$ and $\Omega$, respectively. The expression of $\tilde{f}_{i}, \tilde{g}_{i}$ are
\begin{align}
\tilde{f}_{1} & =\frac{M_{2}}{M_{1}\left(M_{1}^{2}-2M_{1}M_{2}+M_{2}^{2}-q^{2}\right)\left(M_{1}^{2}+2M_{1}M_{2}+M_{2}^{2}-q^{2}\right)^{2}}\nonumber\\
 & \times\Big[M_{1}^{4}({\rm Im}^2C_{3}^{1}-2{\rm Im}^2C_{1}^{1})+4M_{1}^{3}M_{2}({\rm Im}^2C_{4}^{1}-2{\rm Im}^2C_{1}^{1})\nonumber\\
 & -2M_{1}^{2}\left(M_{2}^{2}({\rm Im}^2C_{1}^{1}+2{\rm Im}^2C_{2}^{1}+{\rm Im}^2C_{3}^{1}-4{\rm Im}^2C_{4}^{1})+q^{2}({\rm Im}^2C_{3}^{1}-{\rm Im}^2C_{1}^{1})\right)\nonumber\\
 & +4M_{1}M_{2}{\rm Im}^2C_{4}^{1}\left(M_{2}^{2}-q^{2}\right)+{\rm Im}^2C_{3}^{1}\left(M_{2}^{2}-q^{2}\right)^{2}\Big],\label{eq:f1expr}\\
\tilde{f}_{2} & =\frac{2M_{2}^{2}}{\left(M_{1}^{2}-2M_{1}M_{2}+M_{2}^{2}-q^{2}\right)^{2}\left(M_{1}^{2}+2M_{1}M_{2}+M_{2}^{2}-q^{2}\right)^{2}}\nonumber\\
 & \times\Big[M_{1}^{4}({\rm Im}^2C_{3}^{1}-8{\rm Im}^2C_{1}^{1})+M_{1}^{3}M_{2}(4{\rm Im}^2C_{1}^{1}-2{\rm Im}^2C_{4}^{1})\nonumber\\
 & -2M_{1}^{2}\left(M_{2}^{2}(4{\rm Im}^2C_{1}^{1}-10{\rm Im}^2C_{2}^{1}+{\rm Im}^2C_{3}^{1}+2{\rm Im}^2C_{4}^{1})+q^{2}({\rm Im}^2C_{3}^{1}-4{\rm Im}^2C_{1}^{1})\right)\nonumber\\
 & +2M_{1}M_{2}{\rm Im}^2C_{4}^{1}\left(q^{2}-M_{2}^{2}\right)+{\rm Im}^2C_{3}^{1}\left(M_{2}^{2}-q^{2}\right)^{2}\Big],\\
\tilde{f}_{3} & =\frac{2M_{2}}{M_{1}\left(M_{1}^{2}-2M_{1}M_{2}+M_{2}^{2}-q^{2}\right)^{2}\left(M_{1}^{2}+2M_{1}M_{2}+M_{2}^{2}-q^{2}\right)^{2}}\nonumber\\
 & \times\Big[M_{1}^{6}(2{\rm Im}^2C_{1}^{1}-{\rm Im}^2C_{3}^{1})+M_{1}^{5}M_{2}({\rm Im}^2C_{3}^{1}-{\rm Im}^2C_{4}^{1})\nonumber\\
 & +M_{1}^{4}\left(M_{2}^{2}(8{\rm Im}^2C_{1}^{1}-8{\rm Im}^2C_{2}^{1}+{\rm Im}^2C_{3}^{1}+2{\rm Im}^2C_{4}^{1})-4{\rm Im}^2C_{1}^{1}q^{2}+3{\rm Im}^2C_{3}^{1}q^{2}\right)\nonumber\\
 & -2M_{1}^{3}\left(M_{2}^{3}(2{\rm Im}^2C_{2}^{1}+{\rm Im}^2C_{3}^{1}-3{\rm Im}^2C_{4}^{1})+M_{2}q^{2}({\rm Im}^2C_{3}^{1}-{\rm Im}^2C_{4}^{1})\right)\nonumber\\
 & +M_{1}^{2}\left(M_{2}^{2}-q^{2}\right)\big(M_{2}^{2}(2{\rm Im}^2C_{1}^{1}-8{\rm Im}^2C_{2}^{1}+{\rm Im}^2C_{3}^{1}+2{\rm Im}^2C_{4}^{1})-2{\rm Im}^2C_{1}^{1}q^{2}\nonumber\\
 &+3{\rm Im}^2C_{3}^{1}q^{2}\big)+M_{1}M_{2}\left(M_{2}^{2}-q^{2}\right)^{2}({\rm Im}^2C_{3}^{1}-{\rm Im}^2C_{4}^{1})-{\rm Im}^2C_{3}^{1}\left(M_{2}^{2}-q^{2}\right)^{3}\Big],\\
\tilde{f}_{4} & =\frac{1}{M_{1}^{2}\left(M_{1}^{2}-2M_{1}M_{2}+M_{2}^{2}-q^{2}\right)\left(M_{1}^{2}+2M_{1}M_{2}+M_{2}^{2}-q^{2}\right)}\nonumber\\
 & \times\Big[M_{1}^{4}({\rm Im}^2C_{3}^{1}-2{\rm Im}^2C_{1}^{1})+M_{1}^{3}M_{2}({\rm Im}^2C_{4}^{1}-2{\rm Im}^2C_{1}^{1})\nonumber\\
 & -2M_{1}^{2}\left(M_{2}^{2}({\rm Im}^2C_{1}^{1}-{\rm Im}^2C_{2}^{1}+{\rm Im}^2C_{3}^{1}-{\rm Im}^2C_{4}^{1})+q^{2}({\rm Im}^2C_{3}^{1}-{\rm Im}^2C_{1}^{1})\right)\nonumber\\
 & +M_{1}M_{2}{\rm Im}^2C_{4}^{1}\left(M_{2}^{2}-q^{2}\right)+{\rm Im}^2C_{3}^{1}\left(M_{2}^{2}-q^{2}\right)^{2}\Big],\\
\tilde{g}_{1} & =\frac{M_{2}}{M_{1}\left(M_{1}^{2}-2M_{1}M_{2}+M_{2}^{2}-q^{2}\right)^{2}\left(M_{1}^{2}+2M_{1}M_{2}+M_{2}^{2}-q^{2}\right)}\nonumber\\
 & \times\Big[M_{1}^{4}({\rm Im}^2C_{3}^{5}-2{\rm Im}^2C_{1}^{5})+4M_{1}^{3}M_{2}(2{\rm Im}^2C_{1}^{5}+{\rm Im}^2C_{4}^{5})\nonumber\\
 & -2M_{1}^{2}\left(M_{2}^{2}({\rm Im}^2C_{1}^{5}+2{\rm Im}^2C_{2}^{5}+{\rm Im}^2C_{3}^{5}+4{\rm Im}^2C_{4}^{5})+q^{2}({\rm Im}^2C_{3}^{5}-{\rm Im}^2C_{1}^{5})\right)\nonumber\\
 & +4M_{1}M_{2}{\rm Im}^2C_{4}^{5}\left(M_{2}^{2}-q^{2}\right)+{\rm Im}^2C_{3}^{5}\left(M_{2}^{2}-q^{2}\right)^{2}\Big],\\
\tilde{g}_{2} & =\frac{2M_{2}^{2}}{\left(M_{1}^{2}-2M_{1}M_{2}+M_{2}^{2}-q^{2}\right)^{2}\left(M_{1}^{2}+2M_{1}M_{2}+M_{2}^{2}-q^{2}\right)^{2}}\nonumber\\
 & \times\Big[M_{1}^{4}({\rm Im}^2C_{3}^{5}-8{\rm Im}^2C_{1}^{5})-2M_{1}^{3}M_{2}(2{\rm Im}^2C_{1}^{5}+{\rm Im}^2C_{4}^{5})\nonumber\\
 & -2M_{1}^{2}\left(M_{2}^{2}(4{\rm Im}^2C_{1}^{5}-10{\rm Im}^2C_{2}^{5}+{\rm Im}^2C_{3}^{5}-2{\rm Im}^2C_{4}^{5})+q^{2}({\rm Im}^2C_{3}^{5}-4{\rm Im}^2C_{1}^{5})\right)\nonumber\\
 & +2M_{1}M_{2}{\rm Im}^2C_{4}^{5}\left(q^{2}-M_{2}^{2}\right)+{\rm Im}^2C_{3}^{5}\left(M_{2}^{2}-q^{2}\right)^{2}\Big],\\
\tilde{g}_{3} & =\frac{2}{\left(M_{1}^{2}-2M_{1}M_{2}+M_{2}^{2}-q^{2}\right)^{2}\left(M_{1}^{2}+2M_{1}M_{2}+M_{2}^{2}-q^{2}\right)^{2}}\nonumber\\
 & \times\Big[M_{1}^{6}(2{\rm Im}^2C_{1}^{5}-{\rm Im}^2C_{3}^{5})-M_{1}^{5}M_{2}({\rm Im}^2C_{3}^{5}+{\rm Im}^2C_{4}^{5})\nonumber\\
 & +M_{1}^{4}\left(M_{2}^{2}(8{\rm Im}^2C_{1}^{5}-8{\rm Im}^2C_{2}^{5}+{\rm Im}^2C_{3}^{5}-2{\rm Im}^2C_{4}^{5})-4{\rm Im}^2C_{1}^{5}q^{2}+3{\rm Im}^2C_{3}^{5}q^{2}\right)\nonumber\\
 & +2M_{1}^{3}\left(M_{2}^{3}(2{\rm Im}^2C_{2}^{5}+{\rm Im}^2C_{3}^{5}+3{\rm Im}^2C_{4}^{5})+M_{2}q^{2}({\rm Im}^2C_{3}^{5}+{\rm Im}^2C_{4}^{5})\right)\nonumber\\
 & +M_{1}^{2}\left(M_{2}^{2}-q^{2}\right)\big(M_{2}^{2}(2{\rm Im}^2C_{1}^{5}-8{\rm Im}^2C_{2}^{5}+{\rm Im}^2C_{3}^{5}-2{\rm Im}^2C_{4}^{5})-2{\rm Im}^2C_{1}^{5}q^{2}\nonumber\\
 &+3{\rm Im}^2C_{3}^{5}q^{2}\big)-M_{1}M_{2}\left(M_{2}^{2}-q^{2}\right)^{2}({\rm Im}^2C_{3}^{5}+{\rm Im}^2C_{4}^{5})-{\rm Im}^2C_{3}^{5}\left(M_{2}^{2}-q^{2}\right)^{3}\Big],\\
\tilde{g}_{4} & =\frac{1}{M_{1}^{2}\left(M_{1}^{2}-2M_{1}M_{2}+M_{2}^{2}-q^{2}\right)\left(M_{1}^{2}+2M_{1}M_{2}+M_{2}^{2}-q^{2}\right)}\nonumber\\
 & \times\Big[M_{1}^{4}({\rm Im}^2C_{3}^{5}-2{\rm Im}^2C_{1}^{5})+M_{1}^{3}M_{2}(2{\rm Im}^2C_{1}^{5}+{\rm Im}^2C_{4}^{5})\nonumber\\
 & -2M_{1}^{2}\left(M_{2}^{2}({\rm Im}^2C_{1}^{5}-{\rm Im}^2C_{2}^{5}+{\rm Im}^2C_{3}^{5}+{\rm Im}^2C_{4}^{5})+q^{2}({\rm Im}^2C_{3}^{5}-{\rm Im}^2C_{1}^{5})\right)\nonumber\\
 & +M_{1}M_{2}{\rm Im}^2C_{4}^{5}\left(M_{2}^{2}-q^{2}\right)+{\rm Im}^2C_{3}^{5}\left(M_{2}^{2}-q^{2}\right)^{2}\Big].\label{eq:g4expr}
\end{align}
Here for simplicity we have omitted the superscript $'\rm QCD'$ on the right hand sides. Now, by calculating all the double imaginary parts ${\rm Im}^2 C_{i}^{1,5}$ at the QCD level, we can obtain the analytical expressions for the form factors.

\section{QCD level calculation}\label{sec:QCDLevel}
\begin{figure*}
\begin{center}
\includegraphics[width=0.49\columnwidth]{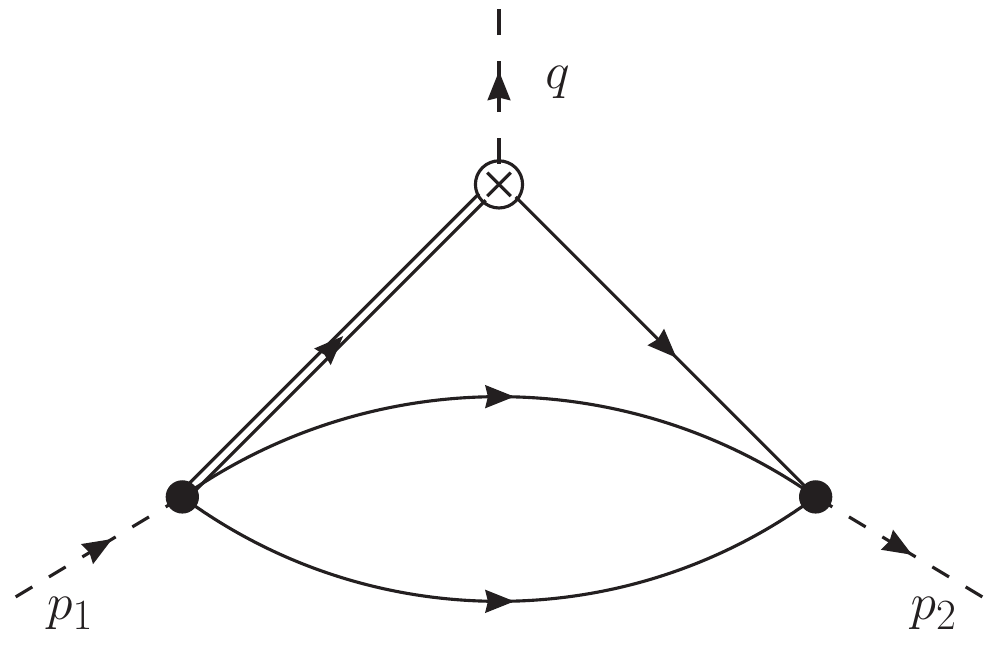} 
\caption{The perturbative diagram for the correlation function. The white crossed dot represents the weak current, and the two black dots denote the interpolating currents for $\Omega_c$ and $\Omega$. The double line corresponds to the charm quark, while the single lines represent the strange quarks.}
\label{sec:PertDiagram} 
\end{center}
\end{figure*}
In this section, we introduce the main approach to calculating the correlation functions $\Pi_{i}(p_2,p_1)$ at the QCD level, formulated as a double dispersion integral. The overall computational methodology follows the approach employed in our previous work Ref.\cite{Shi:2019hbf}. In the deep Euclidean region, where $p_1^2, p_2^2, q^2 \ll 0$, the analytical calculation of $\Pi_{i}(p_2,p_1)$ at  the QCD level  is facilitated through OPE, leveraging the explicit expressions of the three currents provided in Eq.~(\ref{eq:correFunc}). The OPE results are expanded as a series of condensate matrix elements, each associated with progressively higher dimensions. For the purposes of this calculation, the OPE series is truncated to include contributions from operator matrix elements up to dimension 5.

 The leading term in the OPE series corresponds to the condensate of the identity operator, which represents the matrix element of the perturbative diagram illustrated in Fig.~\ref{sec:PertDiagram}, where each line denotes a free quark propagator. The amplitude corresponding to this leading term assumes the form
\begin{align}
\Pi_{i}^{{\rm pert}}(p_{1},p_{2})=\int\frac{d^{4}k_{1}}{(2\pi)^{4}}\frac{d^{4}k_{1}}{(2\pi)^{4}}\frac{N_{i}^{{\rm pert}}(p_{1},p_{2,},k_{1},k_{2})}{(k_{1}^{2}-m_{s}^{2})(k_{2}^{2}-m_{s}^{2})[(p_{1}-k_{1}-k_{2})^{2}-m_{c}^{2}][(p_{2}-k_{1}-k_{2})^{2}-m_{s}^{2}]},
\end{align}
where the nominator $N_{i}^{{\rm pert}}(p_{1},p_{2})$ reads as
\begin{align}
&N_{i}^{{\rm pert}}(p_{1},p_{2},k_1,k_2)\nonumber\\
 =&-2N_{c}!\ {\cal P}^{\alpha\mu}\Big\{(\slashed p_{2}-\slashed k_{1}-\slashed k_{2}+m_{s})\gamma_{\mu}(\slashed p_{1}-\slashed k_{1}-\slashed k_{2}+m_{c})\gamma_{5}\gamma_{\beta}{\rm tr}[(\slashed k_{1}+m_{s})\gamma^{\beta}(\slashed k_{2}-m_{s})\gamma_{\alpha}]\nonumber\\
 & +2(\slashed k_{1}+m_{s})\gamma^{\beta}(\slashed k_{2}-m_{s})\gamma_{\alpha}(\slashed p_{2}-\slashed k_{1}-\slashed k_{2}+m_{s})\gamma_{\mu}(\slashed p_{1}-\slashed k_{1}-\slashed k_{2}+m_{c})\gamma_{5}\gamma_{\beta}\Big\}.
\end{align}
The correlation function must now be expressed in the form of a double dispersion integral, as shown in Eq.~(\ref{eq:QCDPidisper}), where the double imaginary part of the correlation function is extracted. To achieve this, we apply the cutting rules to place each propagator in Fig.~\ref{sec:PertDiagram} on shell, performing the replacement $1/(k^2 - m^2) \to (-2\pi i) \delta(k^2 - m^2)$. This yields
\begin{align}
&{\rm Im}^{2}\Pi_{i}^{{\rm pert}}(p_{1},p_{2}) =\frac{1}{(2i)^{2}}{\rm Disc}^{2}\Pi_{i}^{{\rm pert}}(p_{1},p_{2})\nonumber \\
 =&(-2\pi i)^{4}\int\frac{d^{4}k_{1}}{(2\pi)^{4}}\frac{d^{4}k_{1}}{(2\pi)^{4}}N_{i}^{{\rm pert}}(p_{1},p_{2},k_{1},k_{2})\nonumber \\
 & \times\delta(k_{1}^{2}-m_{s}^{2})\delta(k_{2}^{2}-m_{s}^{2})\delta[(p_{1}-k_{1}-k_{2})^{2}-m_{c}^{2}]\delta[(p_{2}-k_{1}-k_{2})^{2}-m_{s}^{2}]\nonumber \\
 =&-\frac{1}{4(2\pi)^{4}}\int dm_{12}^{2}\int d\Phi_{\Delta}[p_{1},p_{2},m_{c},m_{s},m_{12}]\int d\Phi_{2}[k_{12},m_{s},m_{s}]\ N_{i}^{{\rm pert}}(p_{1},p_{2},k_{1},k_{2}),
\end{align}
where the doubly imaginary part is expressed by a convolution of two phase space integrations. One of them is the standard two-body phase space integration $d\Phi_{2}$, and another one is a triangle diagram integration $d\Phi_{\Delta}$. The corresponding integration measures are
\begin{align}
\int d\Phi_{2}[p,m_{1},m_{2}]&=\int d^{4}k_{1}d^{4}k_{2}\delta^{4}(p-k_{1}-k_{2})\delta(k_{1}^{2}-m_{1}^{2})\delta(k_{2}^{2}-m_{2}^{2}),\nonumber\\
\int d\Phi_{\Delta}[p_{1},p_{2},m_{1},m_{2},m]&=\int d^{4}k\ \delta(k^{2}-m^{2})\delta[(p_{1}-k)^{2}-m_{1}^{2}]\delta[(p_{2}-k)^{2}-m_{2}^{2}].
\end{align}
The scalar integration  of the  triangle diagram can be explicitly calculated as
\begin{align}
\int d\Phi_{\Delta}(p_1,p_2,m_1,m_2,m)\cdot 1=\frac{\pi}{2\sqrt{\lambda}}\Theta[s_1,s_2,q^2,m_1,m_2,m]\theta[s_1-s_{1}^{\rm min}]\theta[s_2-s_{2}^{\rm min}],\label{eq:scalarIntg}
\end{align}
where $\lambda=(s_1+s_2-q^2)^2-4s_1 s_2$ with $p_1^2=s_1, p_2^2=s_2$. $\Theta$ is a $\theta$ function constraining the $s_1, s_2, q^2$:
\begin{align}
&\Theta[s_1,s_2,q^2,m_1,m_2,m]\nonumber\\
=&\theta\big[-m_2^4 s_1 - m_1^4 s_2 + m_2^2 [m^2 (q^2 + s_1 - s_2)+ s_1 (q^2 - s_1 + s_2)]\nonumber\\
 &- 
 q^2 (m^4 + s_1 s_2- m^2 (-q^2 + s_1 + s_2))+ 
 m_1^2 [(q^2 + s_1 - s_2) s_2\nonumber\\
 & + m^2 (q^2 - s_1 + s_2) + m_2^2 (-q^2 + s_1 + s_2)]\big],
\end{align}
The remaining two $\theta$ functions ensure that $s_1$ and $s_2$ lie above the corresponding quark-level thresholds, specifically $s_{1,2} > s_{1,2}^{\rm min} = (m_{c,s} + 2 m_s)^2$. The integration of higher-rank tensors, such as $\int d\Phi_{\Delta} k^{\mu} k^{\nu} \cdots$, can be reduced to scalar integrals of the form given in Eq.~(\ref{eq:scalarIntg}) using the Passarino-Veltman reduction. Detailed results of this reduction can be found in the appendix of Ref.~\cite{Shi:2023kiy}.

\begin{figure*}
\begin{center}
\includegraphics[width=0.9\columnwidth]{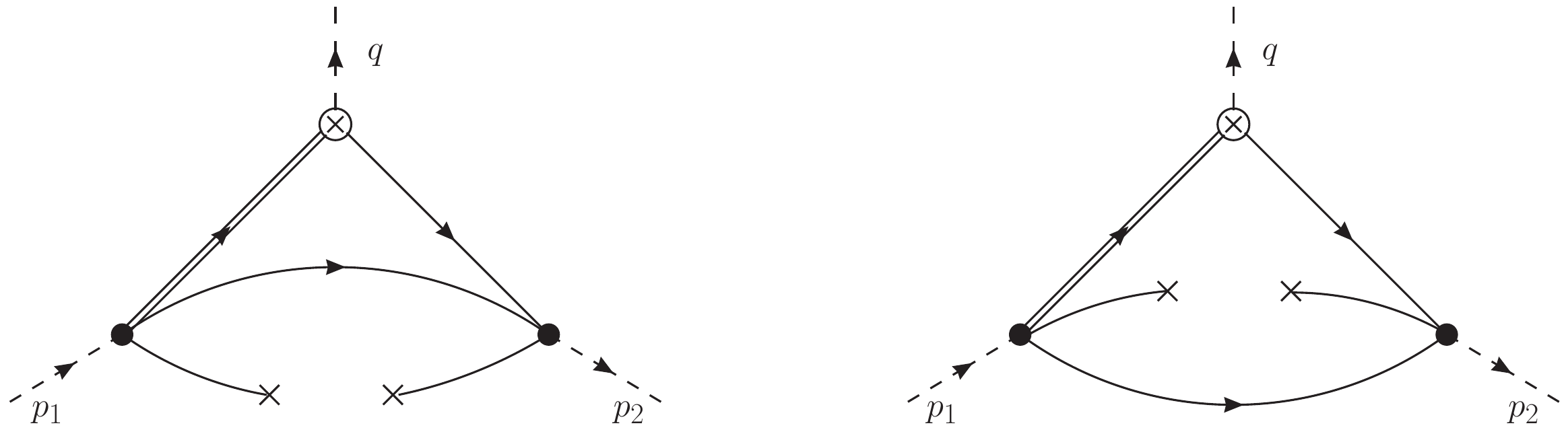} 
\caption{The $\bar q q$ condensate diagram for the correlation function. One of the light quark lines is disconnected, allowing the remaining quark fields represented by the crosses to condense in the vacuum}
\label{sec:qqDiagram} 
\end{center}
\end{figure*}
The next term in the OPE arises from the condensate of the $\bar{q}q$ operator. The corresponding matrix element is illustrated in Fig.~\ref{sec:qqDiagram}, where one of the light quark lines is disconnected, allowing the remaining quark fields represented by the crosses to condense in the vacuum. The $\bar{q}q$ condensate is defined as
\begin{align}
 &\langle 0|{\bar q}_a^i(0) q_b^j(0)|0\rangle= \frac{1}{12}\delta_{ij}\delta_{ba}\langle\bar q q\rangle,\label{eq:nonlocalqq}
\end{align}
where $i,j$ and $a,b$ denote color and spinor indexes, respectively. It should be noted that in Fig.~\ref{sec:qqDiagram}, only the lower two quark lines can be disconnected. If the side strange quark line is disconnected, the diagram would lose its discontinuity in the $p_2$ channel, resulting in the vanishing of the corresponding double imaginary part.

\begin{figure*}
\begin{center}
\includegraphics[width=0.9\columnwidth]{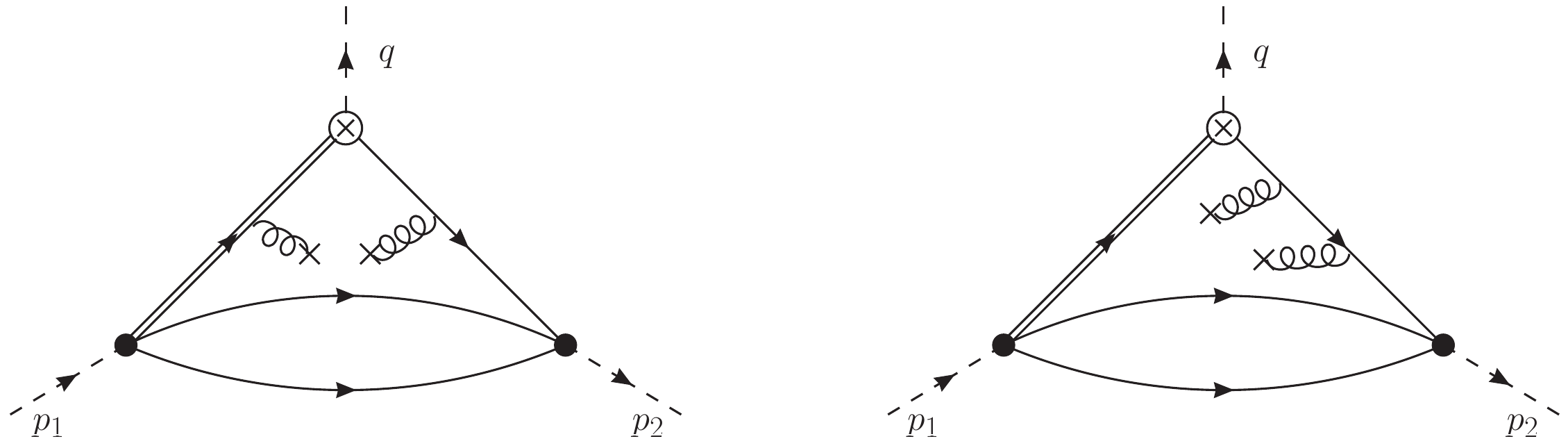} 
\caption{The $GG$ condensate diagram for the correlation function. As an illustration, here two representative diagrams are shown. In  the left diagram, two quark lines interact with a single background gluon field, while in the right diagram, one quark line interacts with two background gluon fields.}
\label{sec:GGDiagram} 
\end{center}
\end{figure*}
The third and fourth terms in the OPE correspond to the contributions from the $GG$ and $\bar{q}Gq$ condensates, respectively. The $GG$ condensate originates from diagrams in which internal quarks interact with two soft gluons. As an illustration, two representative diagrams are shown in Fig.~\ref{sec:GGDiagram}. In  the left diagram, two quark lines interact with a single background gluon field, while in the right diagram, one quark line interacts with two background gluon fields. The quark propagator in the presence of a background gluon field is expressed as
\begin{align}
D(x,0)=&\ i\int \frac{d^4 k}{(2\pi)^4}e^{-i k\cdot x}\left[\frac{\delta_{ij}}{\slashed k -m}-\frac{g_s G_{\alpha\beta}^A t_{ij}^A}{4}\frac{\sigma^{\alpha\beta}(\slashed k+m)+(\slashed k+m)\sigma^{\alpha\beta}}{(k^2-m^2)^2}\right.\nonumber\\
&\left.-\frac{g_s^2(t^A t^B)_{ij}G_{\alpha\beta}^A G_{\mu\nu}^B[f^{\alpha\beta\mu\nu}(k)+f^{\alpha\mu\beta\nu}(k)+f^{\alpha\mu\nu\beta}(k)]}{4(k^2-m^2)^2}+\cdots\right],\label{eq:propagators}
\end{align}
where
\begin{align}
    f^{\alpha\beta\mu\nu}(k)=\  (\slashed k +m)\gamma^{\alpha}(\slashed k +m)\gamma^{\beta}(\slashed k +m)\gamma^{\mu}(\slashed k +m)\gamma^{\nu}(\slashed k +m),
\end{align}
The gluon condensate is defined as
\begin{align}
g_s^2 \langle 0 | G_{\alpha\beta}^A(0) G_{\kappa\tau}^B(0) | 0\rangle=\frac{1}{96}\langle GG \rangle \delta_{AB} (g_{\alpha\kappa}g_{\beta\tau}-g_{\alpha\tau}{\beta\kappa}).
\end{align}
Note that with the use of Eq.~(\ref{eq:propagators}), before extracting the double imaginary part of the correlation function, the $(k^2-m^2)^2$ in the denominator should be reduced to $k^2-m^2$ using the trick: $1/(k^2-m^2)^2=(\partial/\partial M^2)[1/(k^2-M^2)]|_{M^2= m^2}$.
\begin{figure*}
\begin{center}
\includegraphics[width=0.9\columnwidth]{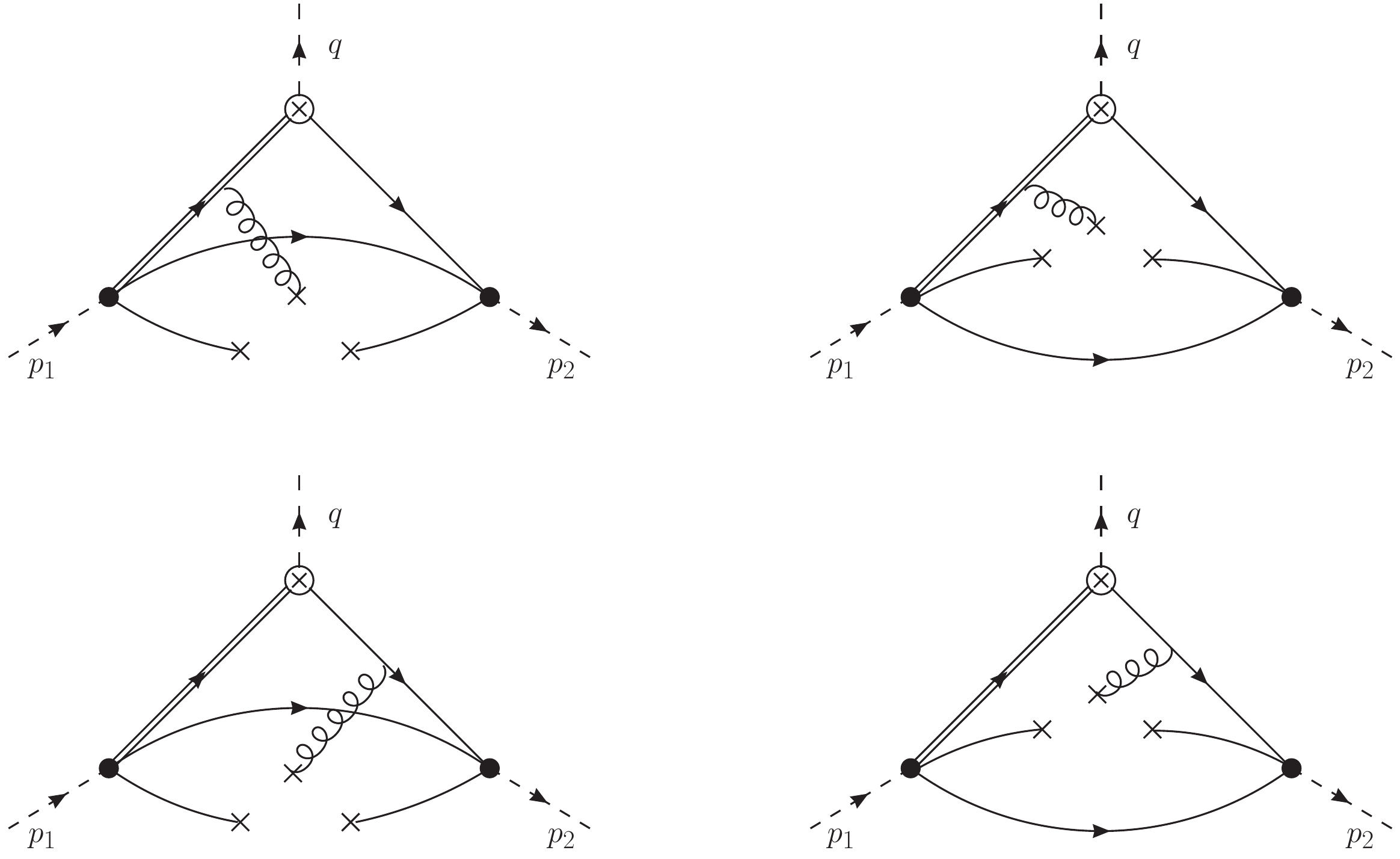} 
\caption{The $\bar q Gq$ condensate diagram for the correlation function. One quark line is disconnected, while the other quark line interacts simultaneously with a background gluon field and two light quark fields}
\label{sec:qGqDiagram} 
\end{center}
\end{figure*}

The $\bar q G q$ condensate comes from the diagrams where one quark line is disconnected, while the other quark line interacts simultaneously with a background gluon field and two light quark fields, as shown in Fig.~\ref{sec:qGqDiagram}. The $\bar q G q$ condensate is defined as
\begin{align}
\langle 0|{\bar q}_a^i(x) g_s G_{\alpha\beta}^A(0)q_b^i(y)|0\rangle=\frac{1}{192}\langle \bar q G q\rangle (\sigma_{\alpha\beta})_{ba} t_{ji}^A,
\end{align}
For the same reason as that of $\bar q q$ condensate, in Fig.~\ref{sec:qGqDiagram} only the lower two quark lines can be disconnected, otherwise the double imaginary part vanishes.

Finally, using the obtained results for the correlation functions at the QCD level and extracting the coefficients $C_i^1, C_i^5$ corresponding to the $\slashed p_2\slashed p_1\gamma_5$ and $\slashed p_2\slashed p_1$ structures, the form factors can be determined through Eqs.~(\ref{eq:f1expr}-\ref{eq:g4expr}).

\section{Numerical results}\label{sec:numericalResult}
In this section, we firstly give the numerical results for the form factors defined in Eq.~(\ref{eq:formfactorDef}). 
In this work, we use the $\overline{\rm MS}$ masses for the quarks, $m_c(\mu)=1.27$~GeV and
$m_s(\mu)=0.103$~GeV with $\mu= 1.27$~GeV \cite{ParticleDataGroup:2020ssz}.  The masses and decay constants of $\Omega_c$ and $\Omega$ are taken as \cite{ParticleDataGroup:2020ssz,Aliev:2016jnp}
\begin{align}
m_{\Omega_{c}} & =2.695~{\rm GeV},~~~~~f_{\Omega_{c}}=0.093~{\rm GeV}^{3}\nonumber\\
m_{\Omega} & =1.672~{\rm GeV},~~~~~f_{\Omega}=0.073~{\rm GeV}^{3}.
\end{align}
The condensate parameters are taken as \cite{Ioffe:2005ym,Colangelo:2000dp}: $\langle\bar{q} q\rangle=-(0.24 \pm 0.01 \mathrm~{\rm GeV})^3,\left\langle\bar{q} G q\right\rangle=m_0^2\langle\bar{q} q\rangle \ {\rm with}\  m_0^2=(0.8 \pm 0.2) \mathrm~{\rm GeV}^2$, and $\langle G G\rangle=(4\pi^2)(0.012 \pm 0.004)~\mathrm{GeV}^4$.

First, we need to determine the threshold parameters $s_{1\rm th}$ and $s_{2\rm th}$.  Empirically, they are often set to be $0.5\,\text{GeV}$ higher than the ground state masses. However, the masses of the first excited states of $\Omega_c$ and $\Omega$ are around $3.0\,\text{GeV}$ and $2.0\,\text{GeV}$, respectively, which are only $0.4\,\text{GeV}$ above the ground states \cite{ParticleDataGroup:2020ssz}. Therefore, this empirical approach is not applicable in this case. In this work, we set $s_{1\rm th}$ and $s_{2\rm th}$ at the midpoint between the ground state masses and the first excited state masses, specifically at approximately $0.2\,\text{GeV}$ above the ground states, and assign an uncertainty range of $\pm 0.1\,\text{GeV}$ around this value, namely
\begin{align}
(m_{\Omega_{c}}+0.1 \text{GeV})^2<&s_{1\rm th}<(m_{\Omega_{c}}+0.3 \text{GeV})^2,\nonumber\\
(m_{\Omega}+0.1 \text{GeV})^2<&s_{2\rm th}<(m_{\Omega}+0.3 \text{GeV})^2.\label{eq:sthregions}
\end{align}

\begin{figure*}
\begin{center}
\includegraphics[width=0.49\columnwidth]{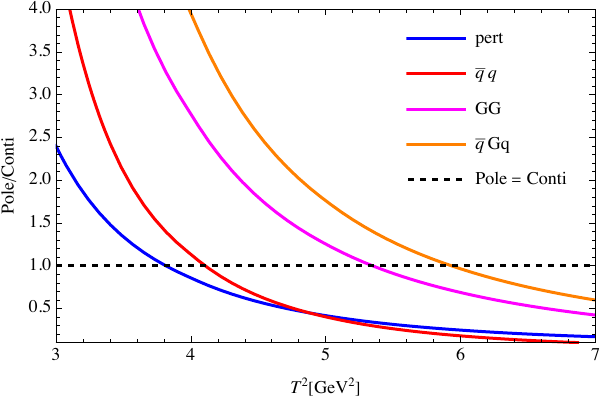} 
\includegraphics[width=0.49\columnwidth]{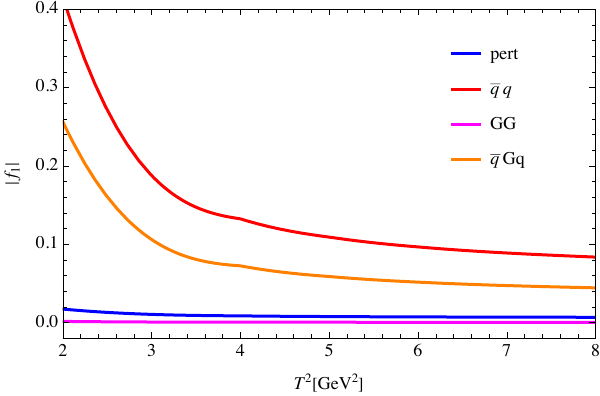} 
\caption{The ${\rm Pole/Conti}$ ratio as a function of $T^2$ for each condensate contribution at $q^2=-8 \text{GeV}^2$ (left); Variation curves of each condensate contribution with respect to $T^2$ (right).}
\label{fig:PoleFracQ2} 
\end{center}
\end{figure*}

After determining the threshold parameters $s_{1,2\rm th}$, one has to select the range for the Borel parameters $T_{1,2}$. To minimize the errors from quark-hadron duality, it is essential to suppress the contributions from the continuum spectrum as much as possible. Therefore, the Borel parameters cannot be too large. We introduce the following criterion to constrain the maximum range of the Borel parameters: the pole contribution must be greater than the contribution from the continuum spectrum. According to Eq.~(\ref{eq:extractFF}), this criterion can be expressed as:
\begin{align}
{\rm Pole/Conti}=\frac{\int_{0}^{s_{1\rm th}}ds_{1}\int_{0}^{s_{2\rm th}}ds_{2}\ e^{-s_1^{2}/T_{1}^{2}}e^{-s_2^{2}/T_{2}^{2}}\left\{ \tilde{f}_{i},\tilde{g}_{i}\right\}}{\int_{s_{1\rm th}}^{\infty}ds_{1}\int_{s_{2\rm th}}^{\infty}ds_{2}\ e^{-s_1^{2}/T_{1}^{2}}e^{-s_2^{2}/T_{2}^{2}}\left\{ \tilde{f}_{i},\tilde{g}_{i}\right\}}>1.\label{eq:PoleConti}
\end{align}
To simplify the problem, the following relation is used to constrain the two Borel parameters corresponding to the $s_1$ and $s_2$ channels \cite{Ball:1991bs}:
\begin{align}
\frac{T_1^{2}}{T_2^{2}} = \frac{m_{\Omega_c}^{2}-m_{c}^{2}}{m_{\Omega}^{2}-m_{s}^2},
\end{align}
and we redefine $T^2=T_1^2$. Here, without loss of generality, we select $f_1$ as an example and choose $q^2$ in the deep Euclidean region, specifically $q^2 = -8\,\text{GeV}^2$, to ensure the convergence of the OPE. The ${\rm Pole/Conti}$ ratio as a function of $T^2$ is plotted separately for each condensate contribution in the left panel of Fig.~\ref{fig:PoleFracQ2}.
It can be observed that for each condensate contribution, the ${\rm Pole/Conti}$ ratio decreases as $T^2$ increases. Based on the criterion ${\rm Pole/Conti} > 1$ and the behavior of the curves for different condensate contributions in the figure, we can determine that the upper limit of $T^2$ lies within the range of $4\ \text{GeV}^2<T_{\rm upper}^2< 6 \text{GeV}^2$.

On the other hand, the Borel parameter must be sufficiently large to ensure the convergence of the OPE. Therefore, to determine the lower limit of the Borel parameter, it is ideally required that the contributions from higher-dimensional condensates are smaller than those from lower-dimensional condensates, that is
\begin{align}
C_{i}^{k,{\rm Pert}}>C_{i}^{k,{\bar q q}}>C_{i}^{k,{GG}}>C_{i}^{k,{\bar q G q}}.\label{eq:hierarchCond}
\end{align}
Taking $f_1$ as an example, the right panel of Fig.~\ref{fig:PoleFracQ2} shows the variation curves of each condensate contribution with respect to $T^2$. However, the figure does not fully exhibit the hierarchical relationship of the condensate contributions as described by Eq.~(\ref{eq:hierarchCond}). As can be seen in the figure, the contributions from the $\bar{q}q$ and $\bar{q}Gq$ condensates are larger than those from the perturbative diagram and the $GG$ condensate. The reason for this is that, compared to the perturbative diagram and the $GG$ condensate, the $\bar{q}q$ and $\bar{q}Gq$ condensate diagrams lack one quark propagator, resulting in the absence of a phase space suppression factor $\int d^4k/(2\pi)^4$. This enhances their contributions. On the other hand, it can be seen from the figure that the contribution from the $\bar{q}q$ condensate is larger than those from the $\bar{q}Gq$ and $GG$ condensates, and the contribution from the perturbative diagram is also larger than that from the $GG$ condensate. This is consistent with the hierarchical relationship given in Eq.~(\ref{eq:hierarchCond}).

From this analysis, it is evident that the lower limit of $T^2$ cannot be strictly determined using Eq.~(\ref{eq:hierarchCond}). Instead, we adopt the stability of the form factor under variations of $T^2$ as an auxiliary criterion for selecting the appropriate range of $T^2$. As illustrated in Fig.~\ref{fig:PoleFracQ2}, the calculation results exhibit increasing stability as $T^2$ increases. While the stability criterion alone would favor larger values of $T^2$, the upper limit for $T^2$ is constrained by the first criterion in Eq.~(\ref{eq:PoleConti}). Consequently, we define the final interval for $T^2$ as $4\ \text{GeV}^2 < T^2 < 6\ \text{GeV}^2$, balancing both stability and the upper bound provided by the ${\rm Pole/Conti}$ ratio.

\begin{figure*}[htp]
\begin{center}
\includegraphics[width=0.96\columnwidth]{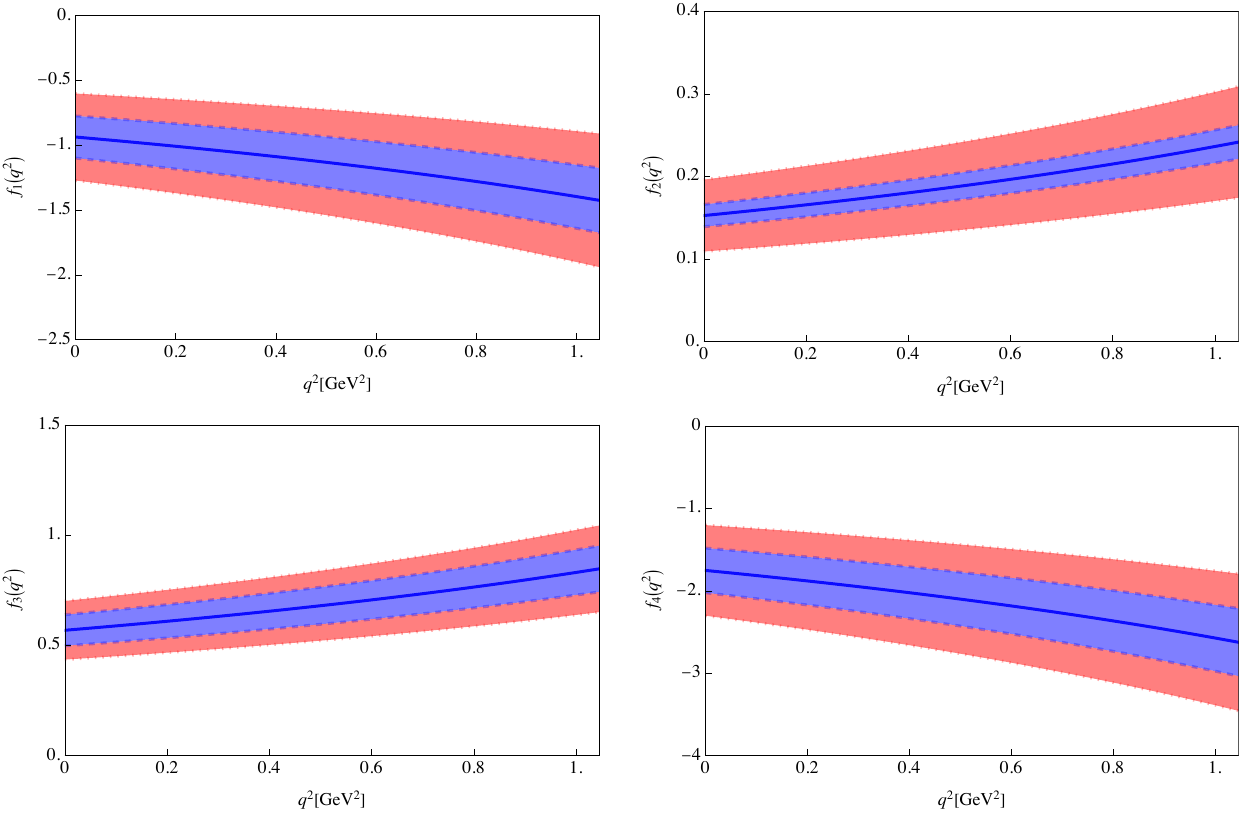} 
\caption{$f_{i}$ as functions of $q^2$ with $s_{1,2\rm th}$ set as Eq.~(\ref{eq:sthregions}) and $4\ \text{GeV}^2<T^2< \text{GeV}^2$. The blue and red bands denote the uncertainty from the choice of $s_{1,2\rm th}$ and $T^2$ in their effective ranges.}
\label{fig:fiplots} 
\end{center}
\end{figure*}

\begin{figure*}[htp]
\begin{center}
\includegraphics[width=0.96\columnwidth]{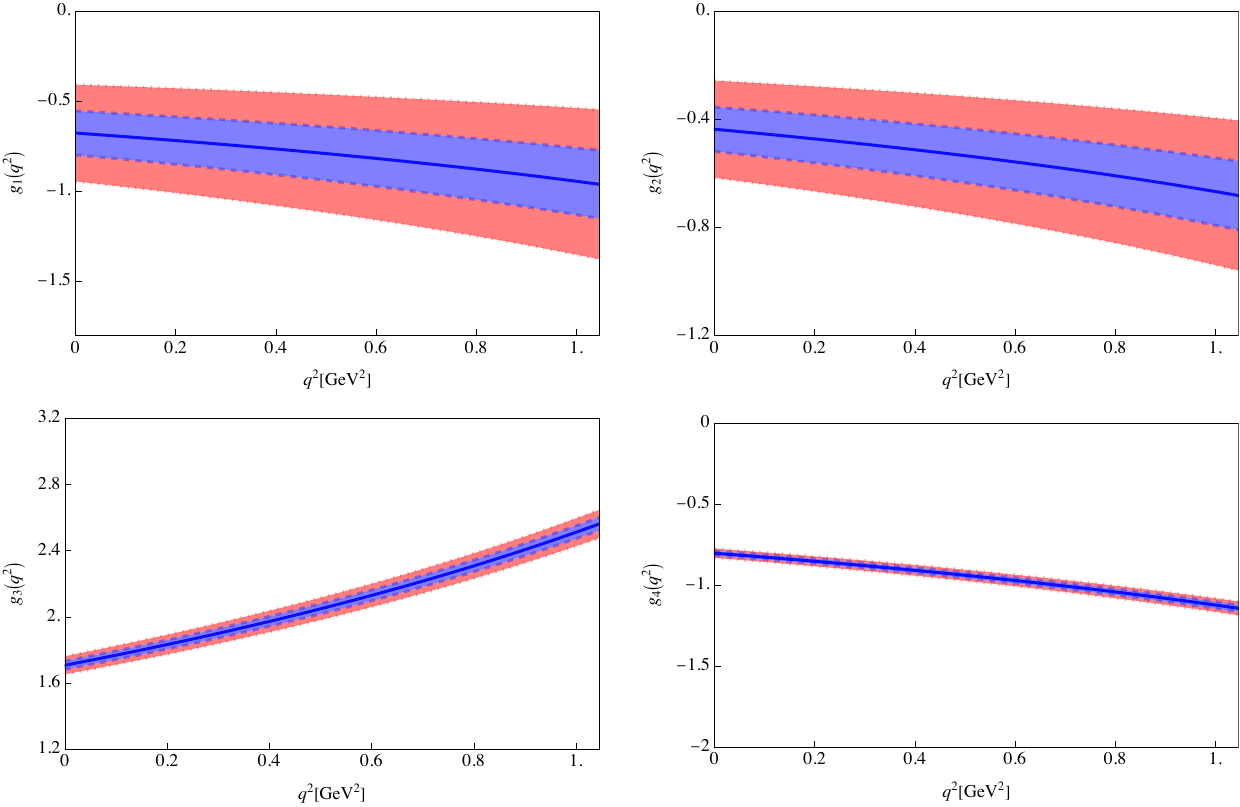} 
\caption{$g_{i}$ as functions of $q^2$ with $s_{1,2\rm th}$ set as Eq.~(\ref{eq:sthregions}) and $4\ \text{GeV}^2<T^2< \text{GeV}^2$. The blue and red bands denote the uncertainty from the choice of $s_{1,2\rm th}$ and $T^2$ in their effective ranges.}
\label{fig:giplots} 
\end{center}
\end{figure*}

The form factors can now be calculated using the chosen values of $s_{1,2\rm th}$ and $T^2$ within their effective regions, as determined above. In the framework of QCDSR, the operator product expansion (OPE) is valid only in the deep Euclidean region. To ensure the convergence of the OPE, the $q^2$-dependence of the form factors is initially computed in the deep Euclidean region, where $q^2 < 0$. By leveraging the analytic properties of the form factors, a fit is performed to the results obtained for $q^2 < 0$, followed by an analytic continuation to the physical region, $q^2 > 0$. The following $z$-expansion fitting function is adopted to describe the form factors within the interval $-10\,\text{GeV}^2 < q^2 < -5\,\text{GeV}^2$:
\begin{align}
F(q^{2})=\frac{1}{1-\frac{q^{2}}{m_{D_{s}}^{2}}}\left[F_{0}+A_{1}\left(z(q^{2})-z(0)\right)\right],\label{eq:FitFunc}
\end{align}
where
\begin{align}
z(t)=\frac{\sqrt{t_{+}-t}-\sqrt{t_{+}-t_{0}}}{\sqrt{t_{+}-t}+\sqrt{t_{+}-t_{0}}},
\end{align}
$t_{\pm}=(m_{\Omega_c}\pm m_{\Omega})^2$, $t_0=t_{+}(1-\sqrt{1-t_{-}/t_{+}})$, $F_0$ and $A_1$ are the two fit parameters. The fit results are listed in Table \ref{Tab:FFfitResult}. In Table \ref{tab:compareFF} the form factors obtained in this work at $q^2=0$ are compared with those calculated using LCSR \cite{Aliev:2022gxi} and LFQM \cite{Hsiao:2020gtc}. From the comparison, it can be observed that the form factors obtained in this work are of the same order of magnitude as those in previous literature.  However, our results are associated with larger uncertainties, primarily arising from the variations in the threshold and Borel parameters. These uncertainties result in more conservative predictions.
\begin{table}
 \caption{The fitting results for the 8 form factors according to Eq.~(\ref{eq:FitFunc}), where the fitting parameters $F_0, a_1$ and their uncertainties are presented.}
\label{Tab:FFfitResult}
\begin{tabular}{|c|cc|c|cc|}
\hline 
$f_i$ & $F_{0}$ & $A_{1}$ &  $g_i$ & $F_{0}$ & $A_{1}$\tabularnewline
\hline 
$f_{1}$ & $-0.93\pm0.33$ & $7.34\pm1.41$ & $g_{1}$ & $-0.67\pm0.26$ & $1.77\pm1.23$\tabularnewline
$f_{2}$ & $0.15\pm0.04$ & $-1.68\pm0.20$ & $g_{2}$ & $-0.43\pm0.16$ & $4.32\pm0.81$\tabularnewline
$f_{3}$ & $0.92\pm0.21$ & $-5.73\pm0.55$ & $g_{3}$ & $2.75\pm0.10$ & $-18.69\pm0.44$\tabularnewline
$f_{4}$ & $-1.75\pm0.54$ & $11.55\pm2.05$ & $g_{4}$ & $-0.80\pm0.02$ & $2.19\pm0.08$\tabularnewline
\hline 
\end{tabular}
\end{table}
\begin{table}
 \caption{Comparing the 8 form factors obtained in this work at $q^2=0$  with those calculated using LCSR \cite{Aliev:2022gxi} and LFQM \cite{Hsiao:2020gtc}. }
 \label{tab:compareFF}
\begin{tabular}{|c|c|c|c|}
\hline 
$F_{0}$ & This Work & LCSR & LFQM\tabularnewline
\hline 
$f_{1}$ & $-0.93\pm0.33$ & $-0.55\pm0.05$ & $0.54\pm0.0$\tabularnewline
$f_{2}$ & $0.15\pm0.04$ & $-0.68\pm0.07$ & $0.35\pm0.01$\tabularnewline
$f_{3}$ & $0.92\pm0.21$ & $1.0\pm0.2$ & $0.33\pm0.02$\tabularnewline
$f_{4}$ & $-1.75\pm0.54$ & $0.16\pm0.02$ & $0.97\pm0.01$\tabularnewline
$g_{1}$ & $-0.67\pm0.26$ & $-0.48\pm0.02$ & $2.05\pm0.05$\tabularnewline
$g_{2}$ & $-0.43\pm0.16$ & $0.68\pm0.07$ & $-0.06\pm0.01$\tabularnewline
$g_{3}$ & $2.75\pm0.10$ & $-1.0\pm0.2$ & $-1.32\pm0.01$\tabularnewline
$g_{4}$ & $-0.80\pm0.02$ & $-0.16\pm0.02$ & $-0.44\pm0.0$\tabularnewline
\hline 
\end{tabular}
\end{table}
In Fig.~\ref{fig:fiplots} and Fig.~\ref{fig:giplots}, we present the form factors as functions of $q^2$ in the deep Euclidean region, where the blue and red bands denote the uncertainty from the choice of $s_{1,2\rm th}$ and $T^2$ in their effective ranges.

The branching fractions of the hadronic and semi-leptonic $\Omega_c\to\Omega$ decays read as \cite{Hsiao:2020gtc}:
\begin{eqnarray}
{\cal B}_h\equiv {\cal B}(\Omega^0_c\to\Omega^- h^+)&=&
\frac{\tau_{\Omega_c}G_F^2|\vec{P}'|}{32\pi m_{\Omega_c}^2}
|V_{cs}V_{ud}^*|^2\,a_1^2 m_h^2 f_h^2 H_h^2\,,\nonumber\\
{\cal B}_\ell\equiv{\cal B}(\Omega_c^0\to\Omega^- \ell^+\nu_\ell)
&=&\frac{\tau_{\Omega_c}G_F^2|V_{cs}|^2}{192\pi^3 m_{\Omega_c}^2}
\int^{(m_{\Omega_c}-m_\Omega)^2}_{m_\ell^2}dq^2
\left(\frac{|\vec{P}'|(q^2-m_\ell^2)^2}{q^2}\right)H_\ell^2\,,
\end{eqnarray}
where $h=\pi, \rho$, and $|\vec{P}'|=\sqrt{Q^2_+ Q^2_-}/(2m_{\Omega})$ with $M_\pm = m_{\Omega_c}\pm m_{\Omega}$, $Q^2_\pm = M_\pm^2 - q^2$. The $H_h, H_l$ read as 
\begin{eqnarray}\label{3H}
H_\pi^2&=&
\left|H_{\frac12{\bar 0}}\right|^2+\left|H_{{-\frac12}{\bar 0}}\right|^2\,,\nonumber\\
H_\rho^2&=&
\left|H_{\frac321}\right|^2+\left|H_{\frac121}\right|^2+\left|H_{\frac120}\right|^2
+\left|H_{-\frac120}\right|^2+\left|H_{-\frac12-1}\right|^2+\left|H_{-\frac32-1}\right|^2\,,\nonumber\\
H_\ell^2&=&\left(1+\frac{m_\ell^2}{2q^2}\right)H_\rho^2+\frac{3m_\ell^2}{2q^2}H_\pi^2\,,
\end{eqnarray}
with $\tau_{\Omega_c}$ the $\Omega_c^0$ lifetime. The Wilson coefficients in Eq.~(\ref{eq:EffHamilton}) are $(c_1,c_2)=(1.26,-0.51)$ at the $m_c$ scale~\cite{Buchalla:1995vs}. The helicity amplitudes given above are defined as $H_{\lambda_1\lambda_2}=H_{\lambda_1\lambda_2}^V-H_{\lambda_1\lambda_2}^A$, and their expressions are
\begin{eqnarray}\label{Hrho}
&& H_{\frac12 {\bar 0}}^{V(A)} =\sqrt{\frac{2}{3}\frac{Q^2_{\pm}}{q^2}}
\left(\frac{Q^2_\mp}{2m_{\Omega_c}m_{\Omega}}\right)
\left[F_1^{V(A)} M_\pm \mp  F_2^{V(A)}\bar M_+ \mp  F_3^{V(A)}\bar M^{\prime}_- \mp F_4^{V(A)} m_{\Omega_c} \right]\,,\\
&&
H_{\frac321}^{V(A)} = \mp \sqrt{Q^2_\mp} \, F_4^{V(A)}\,,\nonumber\\
&&
H_{\frac121}^{V(A)}=-\sqrt{\frac{Q^2_\mp}{3}}
\left[F_1^{V(A)} \left(\frac{Q^2_\pm}{M M'}\right) -F_4^{V(A)}\right]\,,\nonumber\\
&&
H_{\frac120}^{V(A)}= \sqrt{\frac{2}{3}\frac{Q^2_\mp}{q^2}}
\left[ F_1^{V(A)} \left(\frac{Q^2_\pm M_\mp}{2MM'}\right)
\mp\left(F_2^{V(A)}+F_3^{V(A)}\frac{M}{M'}\right) \left(\frac{|\vec{P}'|^2}{M'}\right) 
\mp F_4^{V(A)}\bar M'_- \right]\,,
\end{eqnarray}
where $F_i^V=f_i,\ F_i^A=g_i$, 
and $\bar M_{\pm}=(M_+M_-\pm q^2)/(2m_{\Omega_c})$, $\bar M_{\pm}^{\prime}=(M_+M_-\pm q^2)/(2m_{\Omega})$. 

The obtained branching fractions of hadronic and semi-leptonic $\Omega_c \to \Omega$ decays, as well as their relative ratios to the  benchmark mode $\Omega_{c}^{0}\rightarrow\Omega^{-}\pi^+$ are listed in Table \ref{tab:BrsandCompare}, where a comparison with those from the literatures is also presented.
\begin{table}
\center
\caption{The obtained branching fractions of hadronic and semi-leptonic $\Omega_c \to \Omega$ decays and their relative ratios to the  benchmark mode $\Omega_{c}^{0}\rightarrow\Omega^{-}\pi^+$, compared with those from the literature.} 
\label{tab:BrsandCompare}
\vspace{-6pt}
\setlength{\tabcolsep}{1mm}
\resizebox{\textwidth}{!}{\begin{threeparttable}
\begin{tabular}{cccccccccc}
\hline\hline
\multicolumn{2}{c}{}                   
& This Work &  Fanrong \cite{Zeng:2024yiv} & Cheng
 \cite{Cheng:1996cs} & Wang \cite{Wang:2022zja} &  Gutsche \cite{Gutsche:2018utw}&  Hsiao \cite{Hsiao:2020gtc}  &Aliev \cite{Aliev:2022gxi} &Liu \cite{Liu:2023dvg}\\ \hline
\multicolumn{2}{l}{${\cal B}(\Omega_{c}^{0}\rightarrow\Omega^{-}\pi^+)$}                 
  & $(0.15\pm 0.08)$\% &  $(3.43\pm 0.48)$\%    &  4.19\%  &  1.1\%  & 0.2\%  & 0.51\%& 2.9\% & 1.88\%   \\
\multicolumn{2}{l}{${\cal B}(\Omega_{c}^{0}\rightarrow\Omega^{-}\rho^+)$}                   
& $(7.29\pm 2.10)$\% &   $(18.30\pm 1.94)$\% & 15.08\%   &    & 1.9\% & 1.44$\pm$0.04\% &6.3\%&\\
\multicolumn{2}{l}{${\cal B}(\Omega_{c}^{0}\rightarrow\Omega^{-}e^+\nu_e)$}                   
& $(1.82\pm 0.83)$\% &   $(4.06\pm 0.48)$\% &  &    & & 0.54$\pm$0.02\% &2.06\%&2.54\%\\
\multicolumn{2}{l}{${\cal B}(\Omega_{c}^{0}\rightarrow\Omega^{-}\mu^+\nu_\mu)$}                   
& $(1.75\pm 0.71)$\% &   $(3.81\pm 0.44)$\% &  &    & & 0.50$\pm$0.02\% &1.96\%\\
 \hline
\multicolumn{2}{l}{$R({\Omega^{-}\rho^+})$}                        
& $(46.40\pm 34.32)$ &  $5.33\pm 0.94$  &   3.60  &   & 9.5 &  2.8$\pm$0.4  &2.18&\\
\multicolumn{2}{l}{$R({\Omega^{-}e^+\nu_e})$}                        
& $(11.58\pm 10.45)$ &  $1.18\pm 0.22$&     &   &  &  1.1$\pm$0.2 & 0.71&1.35\\
\multicolumn{2}{l}{$R({\Omega^{-}\mu^+\nu_\mu})$}                        
& $(11.17\pm 9.98)$ &  $1.11\pm 0.20$  &     &   &  &  1.059 & 0.68&\\
\multicolumn{2}{l}{$R({\Omega^{-}e^+\nu_e})/R({\Omega^{-}\mu^+\nu_\mu})$}       &      $(1.037\pm 0.80)$        &  $1.07\pm 0.18$  &     &   &  &  1.08 &1.04 &\\
\hline\hline
\end{tabular}\footnotesize
\end{threeparttable}}
\end{table}
As shown in Table~\ref{tab:BrsandCompare}, the branching fraction of the decay $\Omega_{c}^{0} \rightarrow \Omega^{-}\pi^+$, computed in this work within the framework of QCDSR, is found to be smaller than most theoretical predictions available in the literature. However, it aligns well with the prediction provided by the covariant confined quark model \cite{Gutsche:2018utw}, suggesting a possible consistency between these two approaches. In contrast, the branching fractions for the decays $\Omega_{c}^{0} \rightarrow \Omega^{-}\rho^+$ and $\Omega_{c}^{0} \rightarrow l^+\nu_l$ obtained in this work are in agreement with previous theoretical predictions, being of the same order of magnitude. 

The relatively small branching fraction of $\Omega_{c}^{0} \rightarrow \Omega^{-}\pi^+$ leads to a larger $R$ value when this mode is used as a benchmark, compared to the experimental measurement. This discrepancy may arise from the sensitivity of the $R$ value to the precise determination of the $\Omega_{c}^{0} \rightarrow \Omega^{-}\pi^+$ branching fraction, which is affected by the uncertainties in the form factors. Indeed, the branching fraction predictions presented in this work exhibit considerable uncertainties, primarily due to the significant uncertainties in the form factors. These uncertainties are particularly pronounced in the calculation of the $R$ values, as they propagate through the ratio of branching fractions. Future improvements in the precision of the form factors, either through refined theoretical calculations or additional experimental constraints, will be essential to reduce these uncertainties and provide more reliable predictions.

\section{Summary}\label{sec:conclusion}
In this work, we investigate the weak decays of the charmed baryon $\Omega_c^0$ to the $\Omega^-$ baryon using QCDSR. By defining a three-point correlation function, we calculate the eight form factors governing the $\Omega_c^0 \to \Omega^-$ transition. The cutting rules are employed to extract the double imaginary parts of the correlation functions, enabling their expression as dispersive integrals. These form factors are used to determine the branching fractions for the hadronic decays $\Omega_c^0 \to \Omega^- \pi^+$ and $\Omega_c^0 \to \Omega^- \rho^+$, as well as the semi-leptonic decay $\Omega_c^0 \to \Omega^- \ell^+ \nu_\ell$.
Our results show that the branching fraction of $\Omega_c^0 \to \Omega^- \pi^+$ is smaller than most theoretical predictions but consistent with the covariant confined quark model \cite{Gutsche:2018utw}. In contrast, the branching fractions for $\Omega_c^0 \to \Omega^- \rho^+$ and $\Omega_c^0 \to \Omega^- \ell^+ \nu_\ell$ agree well with previous studies. The relatively small branching fraction of $\Omega_c^0 \to \Omega^- \pi^+$ leads to a larger $R$ value compared to experimental measurements, highlighting the sensitivity of this ratio to uncertainties in the form factors. These uncertainties, arising from variations in the threshold and Borel parameters, are particularly significant in the calculation of the $R$ values.
This study provides a comprehensive analysis of $\Omega_c^0$ decays, offering insights into the dynamics of charmed baryon transitions. The results emphasize the need for precise form factor calculations and suggest directions for future theoretical and experimental improvements.

\section*{Acknowledgement}

We thank Zhen-Xing Zhao for valuable discussions. This work is supported in part by Natural Science Foundation of China under Grants No.12305103, No.12205180.

\end{document}